# Understanding atom probe's analytical performance for iron oxides using correlation histograms and ab initio calculations

Se-Ho Kim[1,2,♛], Shalini Bhatt[1,♛], Daniel K. Schreiber[3], Jörg Neugebauer[1], Christoph Freysoldt[1], Baptiste Gault[1,4,*], Shyam Katnagallu[1,*]

[1] Max-Planck-Institut für Eisenforschung, Max-Planck-Straße 1, 40237 Düsseldorf, Germany.

[2] Materials Science and Engineering, Korea University, Seoul 02841, Republic of Korea.

[3] Energy and Environment Directorate, Pacific Northwest National Laboratory, Richland, WA 99352, United States.

[4] Department of Materials, Imperial College London, Royal School of Mines, Exhibition Road, London, SW7 2AZ, UK.

♛co-first authors; *Corresponding authors: b.gault@mpie.de;  s.katagallu@mpie.de

## Abstract:

Field evaporation from ionic or covalently bonded materials often leads to the emission of molecular ions. The metastability of these molecular ions, particularly under the influence of the intense electrostatic field ($10^{10}$ Vm$^{-1}$), makes them prone to dissociation with or without an exchange of energy amongst them. These processes can affect the analytical performance of atom probe tomography (APT). For instance, neutral species formed through dissociation may not be detected at all or with a time of flight no longer related to their mass, causing their loss from the analysis. Here, we evaluated the changes in the measured composition of $FeO$, $Fe_2O_3$ and $Fe_3O_4$ across a wide range of analysis conditions. Possible dissociation reactions are predicted by density-functional theory (DFT) calculations considering the spin states of the molecules. The energetically favoured reactions are traced on to the multi-hit ion correlation histograms, to confirm their existence within experiments, using an automated Python-based routine. The detected reactions are carefully analysed to reflect upon the influence of these neutrals from dissociation reactions on the performance of APT for analysing iron oxides.

## Introduction:

Atom probe tomography (APT) offers three-dimensional elemental mapping within a small volume of a solid material, achieving sub-nanometer resolution [1],[2]. This method relies on time-controlled ionization and desorption of atoms from a surface, induced by an intense electrostatic field ranging from 1 to 6 × $10^{10}$ Vm$^{-1}$. This process, known as field evaporation [3], is accomplished by biasing a sharp needle-shaped specimen (with a diameter between 20 nm and 200 nm) to a voltage of 2–15 kV. The specimen is maintained at a cryogenic temperature between 20 K and 80 K, optimizing field evaporation conditions and minimizing thermally-activated surface diffusion processes [4],[5].

Originally, atom probe microanalysis focused primarily on metallic materials. However, the introduction of pulsed-laser atom probe[6],[7] has expanded its applicability to poorly conducting materials. Silaeva et al. [8] suggested that semiconductor and insulator surfaces, subjected to intense electric fields, behave similarly to metal surfaces. Despite this, analyzing oxides and nitrides poses challenges due to the emission of intrinsically metastable molecular ions, which tend to dissociate into smaller fragments [9],[10]. The complex energetic paths during dissociation are highly dependent on the intensity of the field [11], [12]. This leads to species specific losses, particularly the anions in oxides and nitrides [13]–[15] which were initially attributed to the direct desorption of neutral species from the specimen's surface (i.e., $N_2/O_2$ gases or N/O neutral atoms) [16], [17]. However, even if the neutral species directly desorb from surface, the post-ionisation probability is too high to explain the compositional biases[15]. These controversies are partly arising from an incomplete understanding of the field evaporation process from these materials.

In the context of the renewed interest in understanding the direct reduction of iron ores by hydrogen gas[18]–[20], particularly leveraging nanoscale characterization techniques like APT for direct reduction [21], by using a H-containing plasma[22], or even using ammonia[23], our study delves into the detailed analyses of $FeO$, $Fe_2O_3$, and $Fe_3O_4$. We systematically varied analysis parameters to study oxygen deficiencies, and to improve the understanding of the formation and dissociation of molecular ions. We complemented experimental observations with density-functional theory calculations for dissociation energies, used jointly with multi-hit ion correlation histograms [9]. These histograms provide valuable insights into correlated evaporation events, molecular dissociations, and the tendency of DC (irrespective of pulsing) evaporation of species in the analyzed material. As such, they allow for retrieving signals that can improve APT's analytical capabilities [15], [24], and provide opportunities for new operational modes, such as analytical field ion microscopy[25]. This work contributes to mapping the analytical capabilities of APT for the analysis of iron oxides, and provides a calibration of compositional measurements, supported by a systematic study of the relative metastability of emitted molecular ions.

## Materials & Methods:

High-grade single crystalline samples of the three oxides were purchased from MaTech Co. APT specimens were prepared from these single crystals with a (0001)-orientation, of hematite and (001)-orientation of magnetite, and wüstite, along the analysis direction (sample z-axis), following the well-established protocol for APT specimen preparation by in-situ lift out by using a scanning- electron microscope (SEM) coupled with a focused-ion beam (FIB) FEI Helios 600 [26]. All specimens were prepared at the same condition (i.e. beam milling current/voltage and time) and targeted for the same radius (see supplementary figure S1). These specimens were then transferred and loaded into a local electrode atom probe (LEAP) 5000 XS system under ambient conditions and analysed at a base temperature of 60 K throughout the measurement process. To investigate the effects of various measurement parameters, we systematically varied pulsed laser energy (10 to 80 pJ), laser frequency (65 kHz to 500 kHz), and detection rates (0.5% to 4%). A total of 1.25 million ions were collected for each analysis condition. For oxidation-state investigation of the hydrogen reduced iron ore, a commercial hematite ore pellet was reduced in a designed furnace using a pure hydrogen gas at flow rate of 30 L/h at 700 °C for 1, 5 and 120 mins. The reduction degree and the reduction rate measured as mass loss per time can be found in the following reference[21].

For analysis, the initial 0.5 million ions were discarded, and only the subsequent 0.75 million ions were utilized to minimize the potential influence from residual heat generated or tip shape changes by previous analysis conditions and delayed field evaporations, and let the specimen adapt to the new conditions of electrostatic field/temperature[27]–[29]. The acquired dataset was then reconstructed to obtain a comprehensive 3D atom map using the CAMECA AP SUITE 6.1 software package.

A unique advantage of straight flight path atom probes is the rich information in correlation histograms for multiple hits of the collected data [9]. The APT datasets contain information regarding the number of ion hits detected in each pulsing window, referred to as multiplicity. For correlation histograms, we focus on events with a multiplicity greater than 1. We preprocess the data to arrange multiple hits with a value greater than two, creating all possible sorted pairs of mass-to-charge values. So, a multiplicity value of 3, corresponding to a detection of triplet ($m_1$, $m_2$, $m_3$) in a pulsing window is processed to 3 sorted pairs {($m_1$, $m_2$), ($m_2$, $m_3$), ($m_1$, $m_3$)} such that the first mass-to-charge value is always smaller than its partner. The resulting mass-to-charge pairs are plotted on a 2D histogram containing two mass-to-charge axes. As discussed by Saxey [9], molecular dissociation often leaves a track on the correlation histogram, allowing us to investigate the type of reactions occurring in the analysis of each oxide. For this purpose, we use DFT to provide a table of possible molecular dissociations in these oxides, and the associated code and notebooks are available on GitHub. The details of the calculations are given below. Using the energies of each molecule in the APT dataset at neutral, + and ++ ionised states, a list of all feasible reactions was created, and these reactions are then tracked on the correlation histograms of each oxide to identify the predicted DFT dissociations.

For the theoretical prediction of dissociation energies, the geometry optimization of neutral, single (+), and double (++) ionized molecules in various spin states was performed with DFT using B3LYP functional and def2-TZVP basis set. Considering different spin states, we aimed to gain insights into the stable electronic states in each case. The final energies of each molecule were determined through a single-point calculation using ORCA 5.0.3 [30]. The starting geometries for some of the molecules were taken from [31], [32]

Results and Discussion:

   A. Calibration:
To ascertain the extent of oxygen loss in each iron oxide sample during field evaporation, we conducted atom probe analyses on hematite (α-$Fe_2O_3$), magnetite ($Fe_3O_4$), and wüstite (FeO). Figure 1 insets, illustrate the reconstructed atom maps. To ensure precise comparisons, three samples were meticulously prepared with specimen geometries closely matching each other, approximately 50 nm in diameter at the tip. In the reconstructed 3D atom maps, iron and oxygen atoms are denoted by pink and cyan dots, respectively.

The stoichiometric fraction in atomic percent of oxygen in hematite, magnetite, and wüstite are 60% (3/5), 57.1% (4/7), and 50% (1/2), respectively, as derived easily from the

oxygen/(oxygen+metal) stoichiometric ratio. A quantitative analysis of the APT data reveals that the observed ratio of oxygen to iron generally decreases as the oxidation state of the iron oxide decreases, as summarized in Figure 1a, but always falls short of the ideal ratio by a factor that sensitively depends on the measurement parameters. During field evaporation, there is a preferential loss of oxygen atoms compared to iron atoms, resulting in a decrease in the O-to-Fe ratio, a phenomenon documented in the analysis of various other oxides and nitrides [13], [33], [34]. The highest oxygen concentrations observed in each sample are approximately 55, 52, and 45 at.%, representing a loss of approximately 5 at.% of oxygen across the three oxides. Notably, there is an overlapping range of oxygen concentrations between 50-52 at.% for hematite and magnetite samples, as well as between 47-43 at.% for magnetite and wüstite.

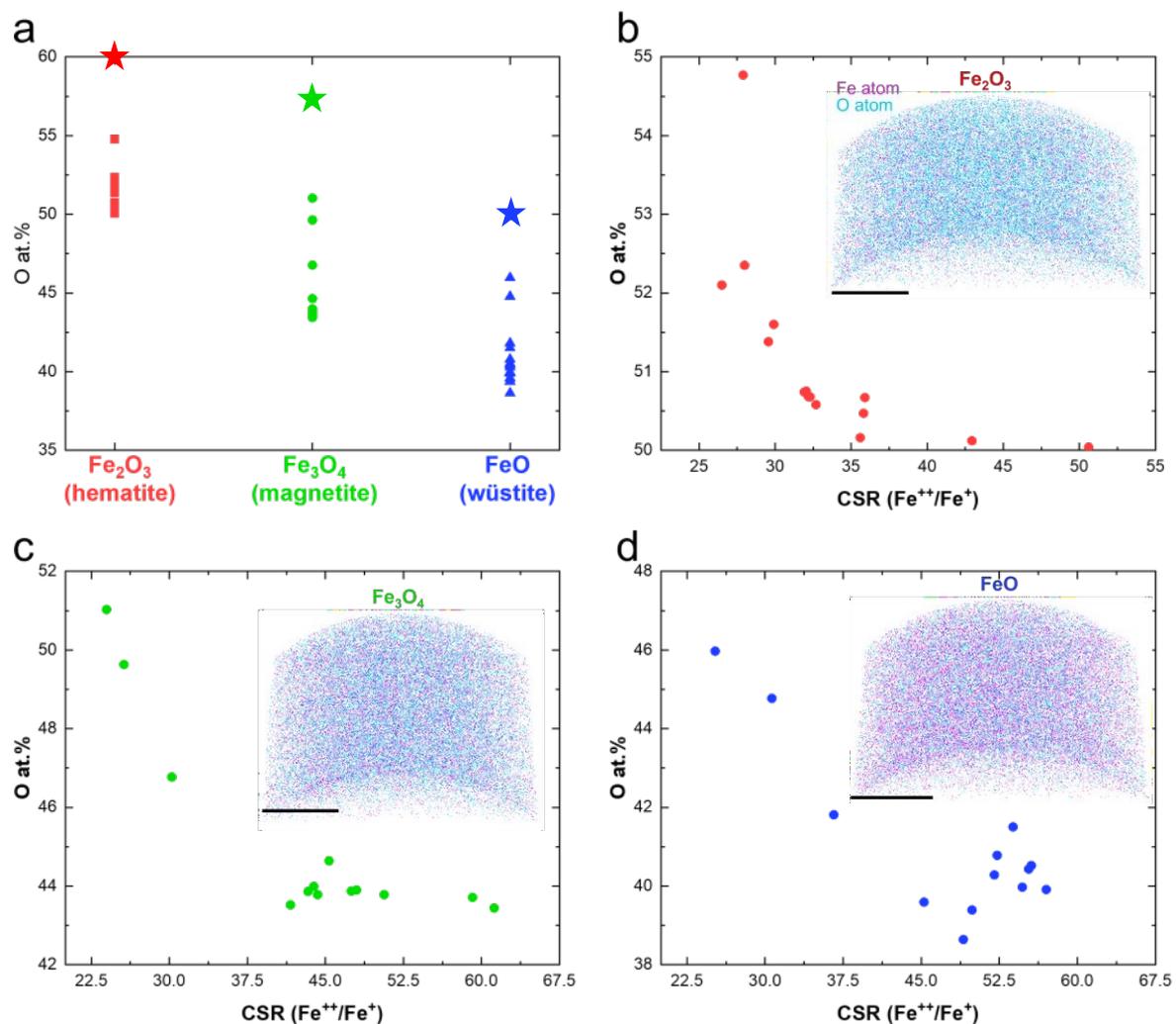

*Figure 1: (a) Summary of elemental O compositions over the variation measurement parameters, with stoichiometric composition of each oxide marked as stars. (b) Evolution of the O composition of (b) hematite, (c) magnetite, (d) wüstite over CSR of Fe++ to Fe+ by varying the measurement parameters (laser energy, laser pulse frequency, detection rate). Insets show the corresponding APT 3D virtual reconstruction volumes for each oxide grown with a (100) orientation, with a scale bar of 10nm.*

To facilitate comparison and enhance the reproducibility of our experiments, we chose to map the experimental parameters by measuring the charge-state ratio (CSR) of Fe. The CSR is defined as the ratio of the number of ions in the peak of $^{56}Fe^+$ to the number of ions in $^{56}Fe^{++}$. While acknowledging its imperfections as a descriptor[35], the CSR serves as an indication of the electrostatic field intensity near the field emitter, with higher laser energy resulting in lower electric fields (e.g., lower CSR). Figure 1b, 1c, and 1d depict the oxygen contents in hematite, magnetite, and wüstite, respectively, plotted against the CSR. Notably, the oxygen contents increasingly deviate from the stoichiometric compositions as the field (and CSR) increases. The observed decay in oxygen with intensity of the electrostatic field is consistent with previous reports [13], [15].

### B. Interpretation

To complement the experimental findings on molecular ion dissociation, we have employed DFT calculations to perform energetic analyses. We considered all experimentally identified (ranged) molecular ion species and theoretically possible combinations. A comprehensive scan over all possible spin configurations was undertaken to identify the most stable spin state of the molecule. Systematically varying spin multiplicities allowed us to determine the relative stability of different spin states. The spin state with the lowest energy was designated as the most stable configuration for the molecule. Detailed results, encompassing formation energies and ionization energies of the most stable spin state, are provided in the supplementary material (Table S1). For the formation energies the chemical potential of Fe, O and H in $Fe_2O_3$, $O_2$ and $H_2$ molecules are chosen as reference. With formation energy then being $E_f = E_{DFT} - n_s \mu_s$, where $n$ and $\mu$ correspond to the number and chemical potential of each species $s$ (Fe, O, H) found in the molecule. Using these energies, we explored all energetically possible dissociation reactions for each molecular species in charge states + and ++. Of course, molecular ions could evaporate in vibrationally or electronically excited states, thereby enabling additional fragmentation pathways. Yet, exploring which excited states are likely to occur - possibly as a function of experimental parameters - is beyond the scope of the present work. Our predicted fragmentation patterns are thus a subset of possible reactions.

Supplementary figure S2 illustrates the energies of 369 dissociation reactions constructed from DFT, with 65 reactions exhibiting negative energies, which indicates the metastability of the parent molecular ion with respect to the daughter species ($m_1$, $m_2$). Table 1 provides a comprehensive compilation of the fragmentation energies of dissociation pathways observed in the experiments and those predicted by DFT to be likely to occur. By analysing the dissociation energies associated with each pathway, we gain valuable insights into the energetics and the relative stability of their different fragmentation products. Across all oxides, two reactions involving the production of neutral $O_2$ are observed experimentally and predicted to have a negative reaction energy from DFT.

|   | Molecular Ion | → | $m_1$ | + | $m_2$ | ΔE (eV) |
|---|---|---|---|---|---|---|
| 1 | $FeO_2^+$ | → | $Fe^+$ | + | $O_2$ | -0.19 |
| 2 | $FeO_2^{++}$ | → | $FeO^+$ | + | $O^+$ | -0.97 |
| 3 | $FeO_2^{++}$ | → | $Fe^+$ | + | $O_2^+$ | -4.92 |
| 4 | $FeO_3^+$ | → | $FeO^+$ | + | $O_2$ | -1.10 |
| 5 | $FeO^{++}$ | → | $Fe^+$ | + | $O^+$ | -0.68 |

| 6 | $Fe_2O_3^{++}$ | → | $FeO_2^+$ | + | $FeO^+$ | -1.21 |
| 7 | $Fe_2O_4^{++}$ | → | $FeO^+$ | + | $FeO_3^+$ | -1.83 |

Table 1: Common reactions predicted by DFT and seen in experimental datasets using correlation histograms. The calculated energy in eV from DFT is given in the last column for each reaction.

In Figure 2a, we illustrate one such reaction $FeO_2^+$ → $Fe^+$ + $O_2$ (reaction 1 in Table 1), with ΔE = -0.19 eV. The trace of this reaction is plotted with a dashed red line on the correlation histogram constructed from the $Fe_2O_3$ dataset. The detection of neutrals is discussed below. Figure 2b showcases another neutral oxygen emitting reaction, $FeO_3^+$ → $FeO^+$ + $O_2$ (reaction 4 in Table 1), with ΔE = -1.1 eV. The traced reactions exhibit a slight deviation from the track observed experimentally. The reaction track aligns correctly if two H are added to the parent molecular ion. However, reaction involving two H is predicted to be positive from DFT. The hydrogen less reaction fits well for $Fe_3O_4$ dataset. One possible explanation is that the thermal pulse provided by the laser creates a surface configuration favorable for the field evaporation of $FeO_2^+$ and $FeO_3^+$. The time taken for the formation of such configurations leads to consistent delayed field evaporation of these ions. This can lead to increased mass-to-charge value for the parent, as evidenced from the experiment, indicating that mass-to-charge adjustment should be done only for the parent ion in both reactions. Further exploration of reasons for the deviation is not undertaken in this study.

For both reactions, the counts along the track are extracted from the histogram and are plotted in Figure 2 (c–d). Both these reactions produce a relatively low cumulative count of 162 and 110, respectively, and this number of neutral species cannot fully explain the loss of oxygen or the underestimation of oxygen from these oxides. However, for neutral species to be detected by the multi-channel plate detector, they must possess a minimum kinetic energy. Research has demonstrated that ions with kinetic energy less than 2 keV fail to generate a sufficiently high electron cascade signal at MCPs to register as an event [36]. Consequently, neutrals produced from the dissociation of the parent molecular species near the tip might not be registered, as the daughter neutral species may not have acquired sufficient kinetic energy for detection. This limitation is evident from the counts extracted from the track, where the cumulative hits reach a plateau (see figure 2c-d).

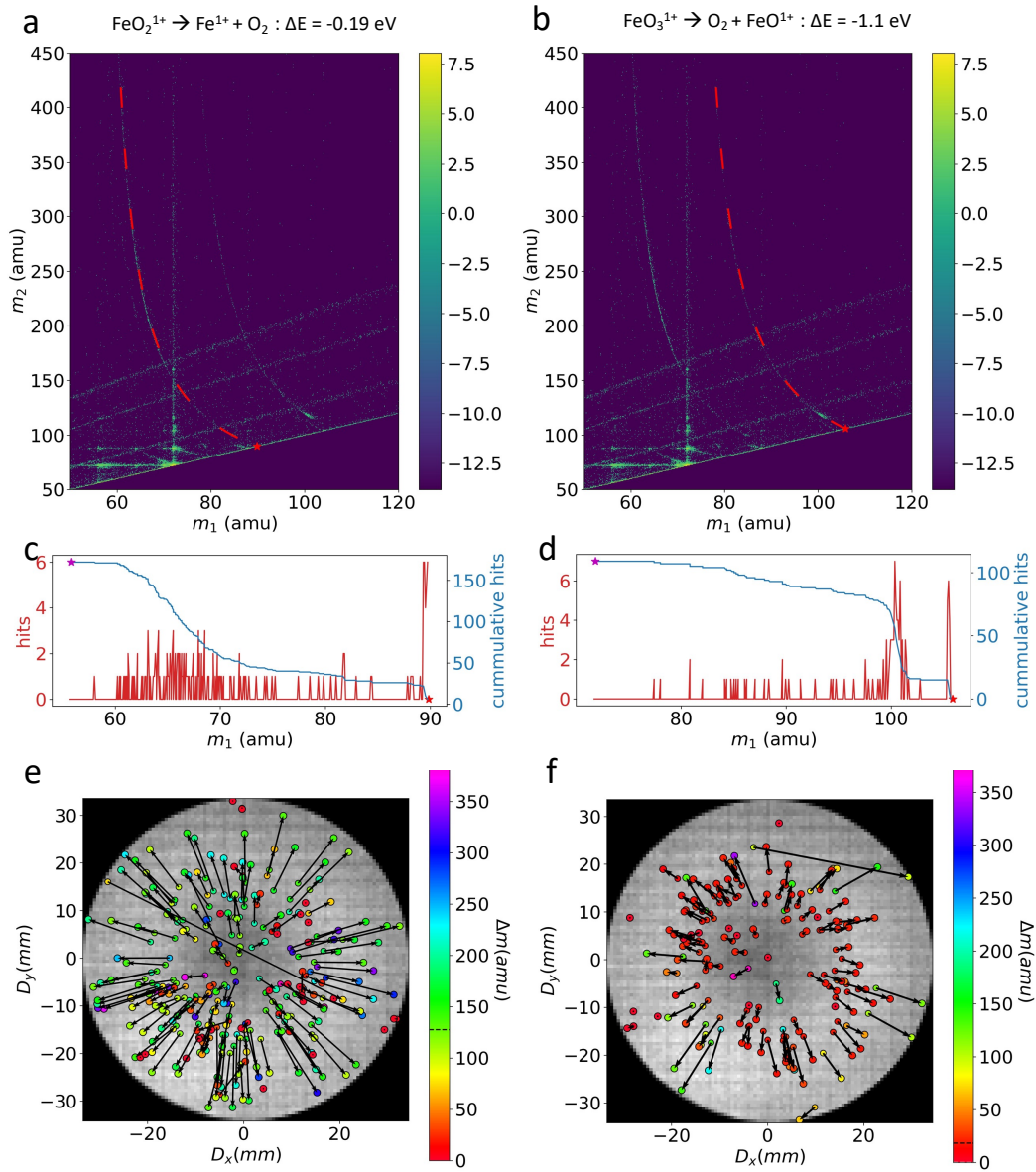

*Figure 2: Correlation histogram with the (a) predicted reaction 1 from table 1 (in dashed red curve) and (b) the predicted reaction 4 (in dashed red curve) traced on in $Fe_2O_3$. (c, d) Counts (red) in the bin along the reaction curve shown in (a, b) along with the cumulative counts(blue) respectively. (e) The detector impact positions of the pairs of ions ($Fe^+$, $O_2$) and (f) the pairs of ions ($FeO^+$, $O_2$) corresponding to the reaction1 and 4. The pairs are related to each other with an arrow point towards, the heavier among the pair, in this case neutral $O_2$. Each pair is colored according to their mass difference, and the median of mass difference is shown in the color bar with a dashed black line. corresponding to the reaction.*

Neutral species may additionally evade detection if their projected impact falls beyond the physical limits of the detector [15]. In a dissociation involving a neutral product, the charged species is consistently closer to the detector center compared to the neutral species. This is attributed to the compression of electric field lines caused by the shank of the specimen. Following dissociation, the neutral species is no longer guided by the field lines and its impact is consistently farther from the center of the detector than the charged species. This is clearly illustrated in Figure 2 (e–f), where the impact positions of pairs corresponding to the reactions are displayed on the detector event histogram. The impact positions of each pair are color-coded based on their mass-to-charge difference, and an arrow is drawn from the lighter of the pair to the heavier ion's impact position. Most arrows for the reactions point outward, indicating that neutral species indeed impact outside the charged species. This is shown schematically in Figure 3. This results in a cutoff in the launch angle of ions from the tip, corresponding to the physical size of the detector, where dissociations leading to a neutral impact position outside the detector are not registered.

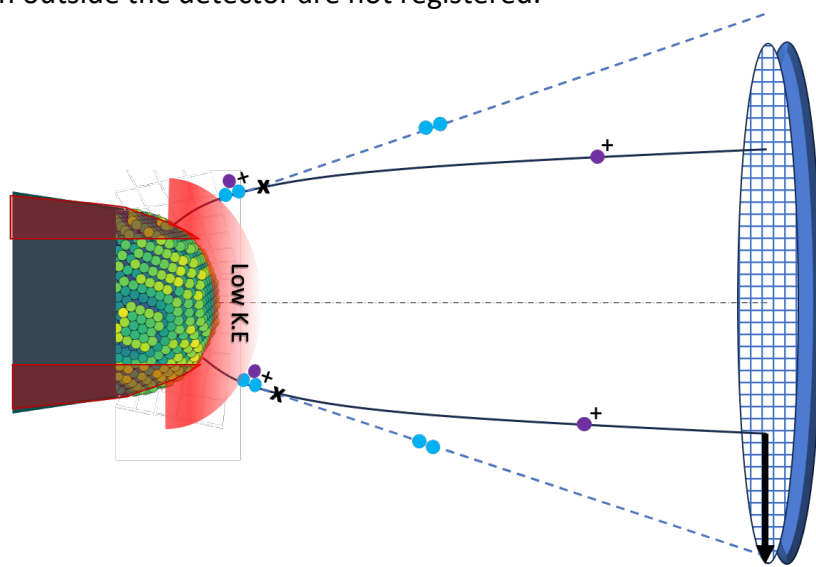

Figure 3: A schematic showing the loss of neutral species (in blue), when dissociation happens from a molecular species field evaporated from dark red region of the tip. When the dissociation is in the unshaded region, the pair of products are detected, with neutral species impacting farther than the charged species, as it can no longer follow the black electric field line. The light red shaded region corresponds to area where a neutral dissociation has little kinetic energy to be detected at the detector.

Lastly, our observations indicate that crystallographic facets play a pivotal role in determining the viability of specific dissociations. Notably, $Fe_2O_3$ possesses a distinct crystal structure compared to FeO and $Fe_3O_4$, resulting in variations in field distribution and the dissociation region on the surface. In Figure 4(a-c), detector event histograms for each oxide are overlaid with the impact positions of dissociation species corresponding to reaction 4. Across all oxides, the impact positions are closely spaced, and the mass-to-charge ratio difference (median ≈ 18 amu) is smaller compared to reaction 1 displayed (median ≈ 130 Da) as seen from Figure 2 (e) and (f). This suggests that dissociation in reaction 4 occurs much closer to the detector across all analysed oxides.

Another distinction lies in the distribution of the dissociation species on the detector, which is contained in a small concentric ring like region for $Fe_2O_3$ and FeO. However, this distribution is relatively homogenous for the $Fe_3O_4$. The detector event histogram (shown as gray-scale background in Figure 4 (a-c)) suggests a low density of hits for $Fe_2O_3$ in the center, indicating a low field region. For $Fe_3O_4$ and FeO, the detector event histograms are more uniform, suggesting a homogenous field distribution. To validate the field distribution, we have also plotted the histograms of events corresponding to H ions (m/n ≤ 3 amu) in Figure 4 (d-e). Total number of ions corresponding to each dataset are also shown. In $Fe_2O_3$, the hotspot of hydrogen impact positions on the detector indicates field inhomogeneity. This suggests an effect of crystallographic direction, leading to an inhomogeneous field or the adsorption of gas species. Consequently, this causes the preferential emission of $FeO_3^+$ from certain crystallographic poles.

We can infer that for these oxides the dissociation reactions leading to neutral species are impacted by the electric field, the crystallographic orientation of the tip and possibly the impact of thermal profile (for creating certain configurations, as indicated by consistent increase in m/n of parent molecular species). Further insight was gained by analyzing mass spectra obtained from the low-density (inner, low field) and high-density (outer, high field) regions of the detector event histogram of $Fe_2O_3$. The mass spectra, depicted in supplementary figure S4, reveal that the high-field region mass spectrum includes additional peaks corresponding to heavier molecular ionic species such as $Fe_2O_3^+$, $Fe_2O_2^+$ and $FeO_4^+$. Our DFT calculations predict that some of these molecular species should dissociate, as shown in supplementary Table S2. The dissociation products then contribute to the molecular species involved in further dissociating into neutral species, as illustrated in Figure 2.

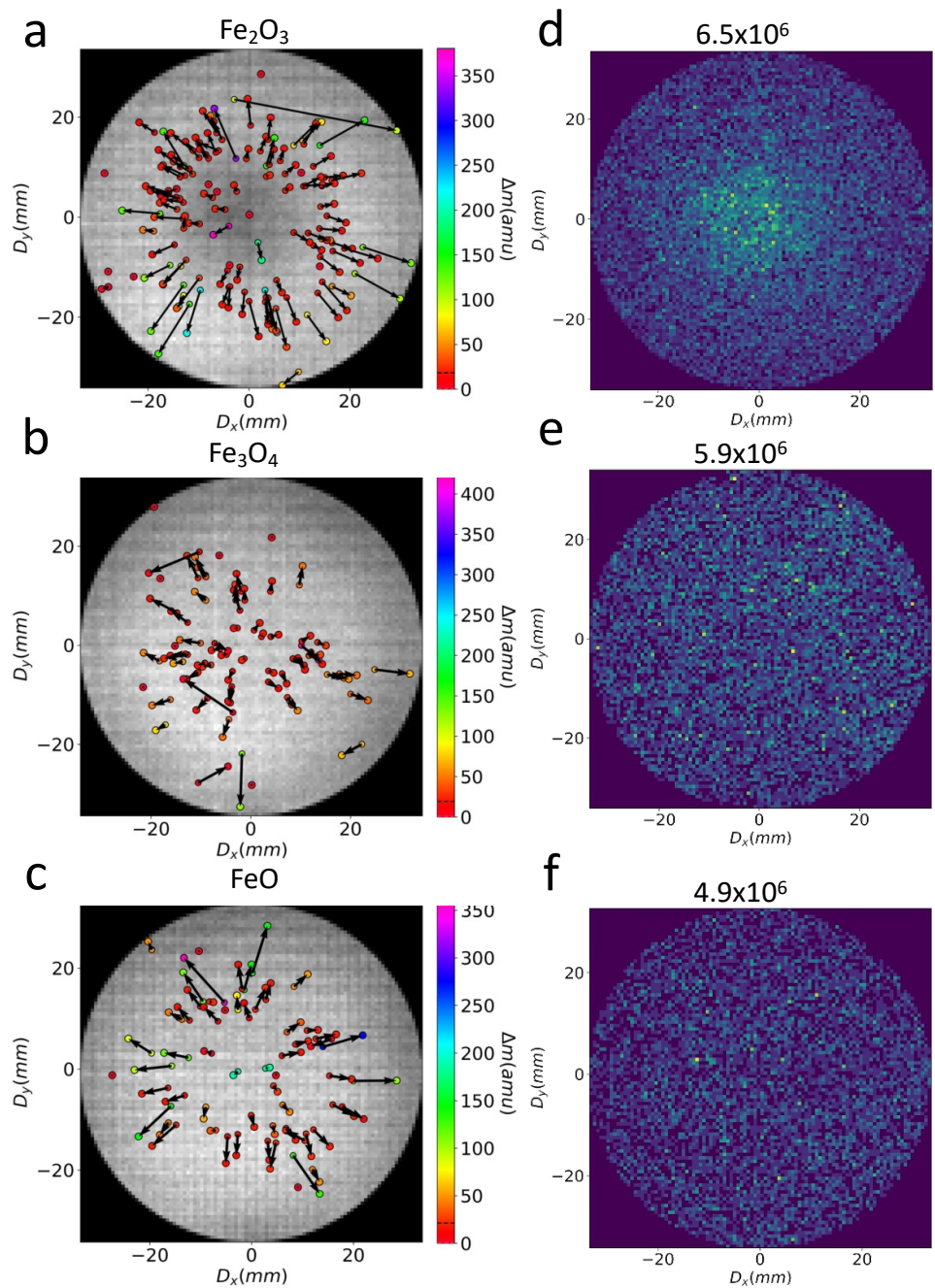

*Figure 4: (a, b, c) Detector event histogram of each oxide overlaid with dissociation products of reaction ($FeO_3^+$ -> $FeO^+$ + $O_2$). (d, e, f) The corresponding Hydrogen detector event histogram of each oxide is shown.*

Beyond dissociations resulting a neutral species, we also observed dissociation reactions into two charged, oxygen-containing species. These reactions are overlaid on each oxide's correlation histogram, as shown in supplementary figure S3. Dissociations involving charged species manifest as curved lines with negative slopes on the correlation plot. One endpoint lies on the diagonal at the mass-to-charge ratio of the parent ion and corresponds to a late dissociation. The other endpoint occurs if the ion splits early on and yields coordinates corresponding to the fragmentation products. Since the reaction species in between these endpoints end up in the tails of the 1D mass spectrum, they are not accurately ranged. All the correlation histograms also exhibit strong correlation events during the tail of laser pulsing, leading to diagonal lines emanating from hotspots of corresponding correlated mass-to-charge ratios, which are also not correctly ranged. The fraction of multiple events in these oxides is significant, ~55% in FeO, ~51% in $Fe_2O_3$, and ~54% in $Fe_3O_4$. Such a high fraction of multiples can introduce compositional biases due to improper ranging.

### C. APPLICATION TO THE ANALYSIS OF DIRECT REDUCTION OF IRON OXIDES

Understanding the kinetics of iron reduction using hydrogen as an alternative reductant has been initiated by analyzing the microstructure and local chemistry of the ores at atomic-scale levels, such as through APT. During this process, the reduction proceeds from iron-ore hematite ($Fe_2O_3$) to magnetite ($Fe_3O_4$), then to wüstite (FeO), and finally to metallic iron (Fe). The mass loss due to the FeO to Fe transition is significant, and the sluggish reduction of the FeO becomes the dominant process. The final reduction step consistently generates non-stoichiometric oxides[37]. Revealing the effect of the presence of these pseudo-phases on the most sluggish step (i.e. wüstite reduction) is crucial for a thorough understanding of the reduction kinetics. However, this is rarely reported due to the detection resolution limits in most studies, which were conducted at the macro- and/or micro-scale rather at the near-atomic-scale[21]. Therefore, precise atomic compositional information of O must be provided to identify which stable/unstable oxides are present. However, a challenge arises as O depletion occurs during field evaporation of oxides; neutral O atoms do not contribute to the detector signal leading to apparent O loss, which means measurements of FeO phases could be mis-interpreted as meta-stable $Fe_{1-x}O$.

As described above, the determination of iron ores reduction from APT can be significantly affected by the measurement parameters of APT. For instance, considering CSR of Fe combined with the simulation results, it could provide insights into the field strength and the corresponding reduction state. As-received commercial iron ores were subjected to reduction using pure hydrogen gas in a custom-designed furnace at 700 °C for durations of 1, 5, and 120 mins (see the Methods). The pristine and H-reduced sample were prepared into APT specimen, with a focus on targeting the surface sites during specimen preparation, and subsequently measured with APT. After 120 min of reduction, the ores fully transformed into iron (O < 0.07 at.%). However, after the 1-min hydrogen reduction process of the commercial iron ore, the O level dropped to approximately 45%. It remains unclear whether this decrease resulted from partial reduction or the influence of laser energy. Additionally, following the 5-minute hydrogen reduction process, two distinct phases with different O contents were detected. The first phase from the 5-minute reduced sample contains approx. 45 at.% O while the second phase contains approx. 37 at.% O. However, solely based on the O ion counts from the acquired data, both the first and second phases might be misinterpreted as pseudophases of wüstite, since their correspondence cannot not be definitively determined without considering the fields that influence the O measurement.

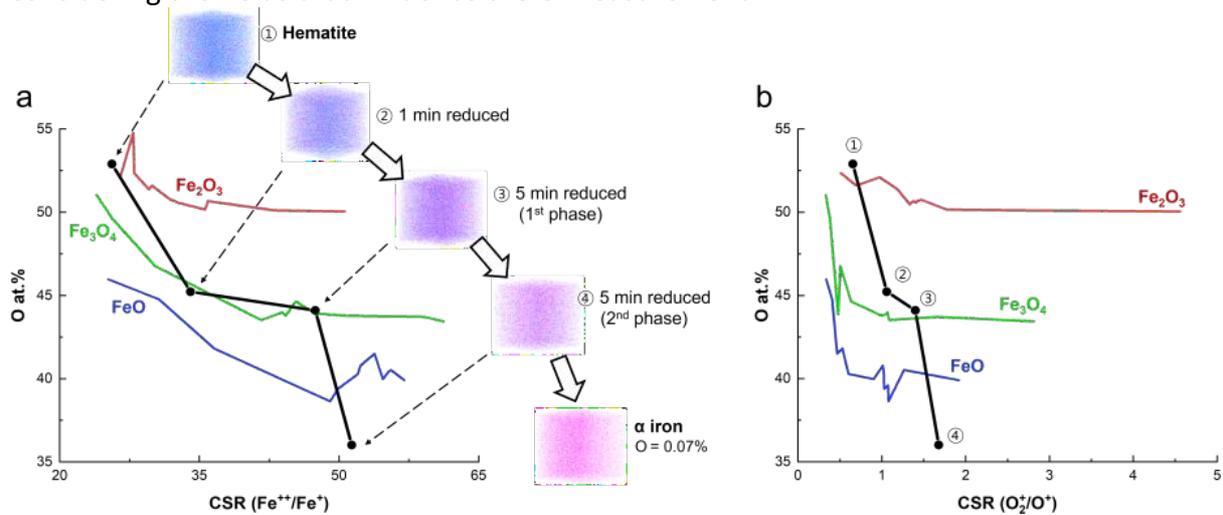

*Figure 5: O concentration vs. CSR of (a) $Fe^{++}/Fe^{+}$ and (b) $O^{++}/O^{+}$ of the H-reduced iron ores (0, 1, 5, 120 mins) and the reference samples of (0001)-grown hematite (red), magnetite (green), and wüstite (blue). Insets show the reconstructed cuboidal maps of the H-reduced samples. Pink and cyan dots represent Fe and O atoms.*

Although the specimens were measured with the same laser pulse energy, each specimen experienced different field conditions depending on their geometry and corresponding field factor [38], [39], leading to different CSR. The O concentrations versus $Fe^{++}/Fe^{+}$ CSR from each H-reduced sample were plotted along with the references of the hematite, magnetite and wüstite sample (see Figure 5a). The as-received sample (iron ores) aligned with the hematite (red line) while the 1-min reduced sample (marked as 2) was closer to the magnetite regime (green). The two different phases (marked as 3 and 4) detected in the 5-mins reduced sample corresponded to magnetite and a lower wüstite regime, suggesting that the second phase is a partially reduced wüstite phase. From our previous discussion, we observe significant differences in the neutral dissociation reactions of hematite when compared to magnetite and wüstite, which behave similarly. This can be ascribed to different crystal structure and electrostatic field conditions. This gives us more confidence to associate the third phase to partially reduced wüstite. Similar trends were observed when plotting the $O^{+}/O_2^{+}$ CSR against O level in Figure 5b [40], [41].

A comprehensive investigation of the various measurement parameters with theoretical dissociation reactions is imperative to guarantee measurement precision in APT analyses of oxide materials. This empirical methodology is vital for both advancing the understanding of APT and its application in material characterization. A meticulous assessment of relevant factors, such as laser pulse energy (e.g., field strength), is essential to ensure the acquisition of reliable and meaningful concentration maps, which is a prerequisite for the analysis of new materials.

## Conclusions

In conclusion, this study underscores the critical need for comprehensive investigations to elucidate the impact of various process parameters on the precision of concentration maps derived from atom probe measurements. By delving into the dissociation reactions through energy calculations, considering diverse spin states, we underscore the pivotal role of stable spin states in predicting these reactions. The utilization of correlation histograms and DFT calculations has enabled us to identify specific dissociation reactions that are consistently observed across all examined oxides. Notably, these dissociations, particularly those involving neutral species, exhibit a crystallographic dependence. Moreover, certain reactions may result in neutral species remaining undetected by the detector due to impact positions lying outside the detection range or possessing insufficient kinetic energy. The substantial fraction of multiple events, characterized by significant correlations, introduces the potential for compositional biases, emphasizing the necessity for meticulous scrutiny of 1D mass spectra in tandem with correlation histograms. This study thus stresses the intricate interplay of various factors in atom probe measurements for these oxides, emphasizing the importance of nuanced analyses for accurate interpretation and reliable results.

### Acknowledgements:
S-HK and BG acknowledge financial support from the German Research Foundation (DFG) through DIP Project No. 450800666. S-HK acknowledges the KIAT grant funded by the Korea Government MOTIE (P0023676). SB gratefully acknowledges financial support from the International Max Planck Research School for Sustainable Metallurgy (IMPRS SusMet). BG is grateful for financial support from the ERC-CoG-SHINE-771602. BG & SK are grateful for funding from the DFG through the Leibniz Prize 2020. DKS acknowledges support from the U.S. Department of Energy (DOE), Office of Science, Basic Energy Sciences, Materials Science & Engineering Division, Mechanical Behavior and Radiation Effects program through FWP 56909 at Pacific Northwest National Laboratory (PNNL) for technical contributions on APT data interpretation. PNNL is a multiprogram national laboratory operated by Battelle for the U.S. DOE under Contract DE-AC05-79RL01830.

### Code Availability:

The related code and notebooks for analysis can be found at the following github page
https://github.com/skatnagallu/CorrelationHistogram

## Supplementary material:

| Molecule | $E_f^0$ (eV) | Spin state | $E_f^1$ (eV) | Spin state | $E_f^2$ (eV) | Spin state | IE(I) (eV) | IE(II) (eV) |
|---|---|---|---|---|---|---|---|---|
| $Fe_2O_2$ | 5.23 | 9 | 12.27 | 10 | 25.09 | 9 | 7.04 | 12.82 |
| $FeO_2$ | 1.27 | 3 | 11.81 | 4 | 28.96 | 11 | 10.54 | 17.15 |
| $Fe_2O$ | 3.91 | 9 | 10.37 | 8 | 25.81 | 11 | 6.46 | 15.44 |
| $FeO_3$ | 0.95 | 3 | 12.42 | 2 | 31.20 | 5 | 11.47 | 18.78 |
| FeOH | 2.17 | 4 | 9.57 | 5 | 26.38 | 6 | 7.40 | 16.81 |
| FeH | 5.53 | 4 | 12.94 | 5 | 30.48 | 4 | 7.40 | 17.54 |
| $HO_2$ | 1.31 | 2 | 12.58 | 3 | 31.30 | 2 | 11.27 | 18.71 |
| $FeO_4$ | 3.06 | 9 | 14.73 | 10 | 32.72 | 11 | 11.66 | 17.99 |
| OH | 1.65 | 2 | 14.73 | 3 | 35.47 | 4 | 13.08 | 20.74 |
| $OH_3$ | 4.43 | 2 | 9.74 | 3 | 33.29 | 3 | 5.31 | 23.55 |
| $H_2O$ | ref | 1 | 12.49 | 2 | 37.53 | 5 | 12.49 | 25.05 |
| O | 2.67 | 3 | 16.67 | 4 | 51.88 | 3 | 14.00 | 35.21 |
| Fe | 3.89 | 5 | 11.62 | 4 | 28.19 | 5 | 7.72 | 16.57 |
| $O_2$ | ref | 3 | 12.43 | 2 | 34.95 | 7 | 12.43 | 22.53 |
| H | 3.59 | 2 | -- | -- | -- | -- | -- | -- |
| $Fe_2$ | 7.69 | 9 | 13.87 | 8 | 26.93 | 9 | 6.14 | 13.10 |
| FeO | 2.33 | 5 | 11.32 | 4 | 28.97 | 3 | 8.99 | 17.64 |
| $H_2$ | 2.40 | 1 | 17.83 | 2 | -- | -- | 15.43 | -- |
| $Fe_2O_3$ | ref | 9 | 8.91 | 8 | 24.34 | 11 | 8.91 | 15.44 |
| $FeO_2H$ | 0.37 | 6 | 10.26 | 5 | 27.48 | 8 | 9.88 | 17.22 |
| $Fe_2O_4$ | -0.75 | 7 | 8.96 | 2 | 25.57 | 7 | 9.71 | 16.61 |
| $H_3$ | 6.16 | 2 | 15.01 | 1 | 37.28 | 2 | 8.85 | 22.27 |
| $FeO_2H_2$ | -0.76 | 5 | 8.24 | 6 | 25.32 | 7 | 9.00 | 17.08 |
| $H_2O_3$ | 1.59 | 1 | 11.70 | 2 | 31.41 | 1 | 10.11 | 19.70 |
| $FeH_2O$ | 3.70 | 5 | 9.89 | 4 | 24.16 | 5 | 6.18 | 14.28 |
| $FeH_2$ | 6.31 | 5 | 15.89 | 6 | 33.23 | 7 | 9.58 | 17.34 |
| $HO_3$ | 1.86 | 2 | 13.04 | 3 | 27.71 | 2 | 11.18 | 14.67 |
| $H_2O_2$ | 1.26 | 1 | 11.69 | 2 | 33.74 | 1 | 10.42 | 22.06 |

| | | | | | | | |
|---|---|---|---|---|---|---|---|
| O₃ | 2.37 | 5 | 14.95 | 2 | 37.36 | 5 | 12.57 | 22.41 |
| FeO₃H₂ | -0.65 | 5 | 9.47 | 4 | 26.89 | 3 | 10.12 | 17.42 |

Table S 1: Formation energies (eV) with respect to chemical potential of Fe, O, H in $Fe_2O_3$, $O_2$ and H2O molecules, and Ionization energies (in eV) of the molecules under the most stable spin state are shown in the table.

| | $m_p$ | charge | $m_1$ | charge | $m_2$ | charge | ΔE |
|---|---|---|---|---|---|---|---|
| 1 | Fe₂O₂ | 2 | FeO | 1 | FeO | 1 | -2.44 |
| 2 | Fe₂O₂ | 2 | FeO₂ | 1 | Fe | 1 | -1.66 |
| 3 | FeO₄ | 1 | O₂ | 0 | FeO₂ | 1 | -2.92 |
| 4 | FeO₄ | 1 | O₂ | 1 | FeO₂ | 0 | -1.03 |
| 5 | FeO₄ | 1 | O₃ | 0 | FeO | 1 | -1.03 |
| 6 | FeO₄ | 2 | O₂ | 0 | FeO₂ | 2 | -3.76 |
| 7 | FeO₄ | 2 | O₂ | 1 | FeO₂ | 1 | -8.49 |
| 8 | FeO₄ | 2 | O₃ | 0 | FeO | 2 | -1.38 |
| 9 | FeO₄ | 2 | O₃ | 1 | FeO | 1 | -6.45 |
| 10 | FeO₄ | 2 | O | 1 | FeO₃ | 1 | -3.63 |
| 11 | Fe₂O₃ | 2 | O₂ | 1 | Fe₂O | 1 | -1.54 |
| 12 | Fe₂O₃ | 2 | FeO₂ | 1 | FeO | 1 | -1.21 |
| 13 | Fe₂O₃ | 2 | Fe | 1 | FeO₃ | 1 | -0.31 |

Table S 2: Reactions predicted by DFT for heavier molecular species seen in high field mass spectra of $Fe_2O_3$. The calculated energy in eV from DFT is given in the last column for each reaction.

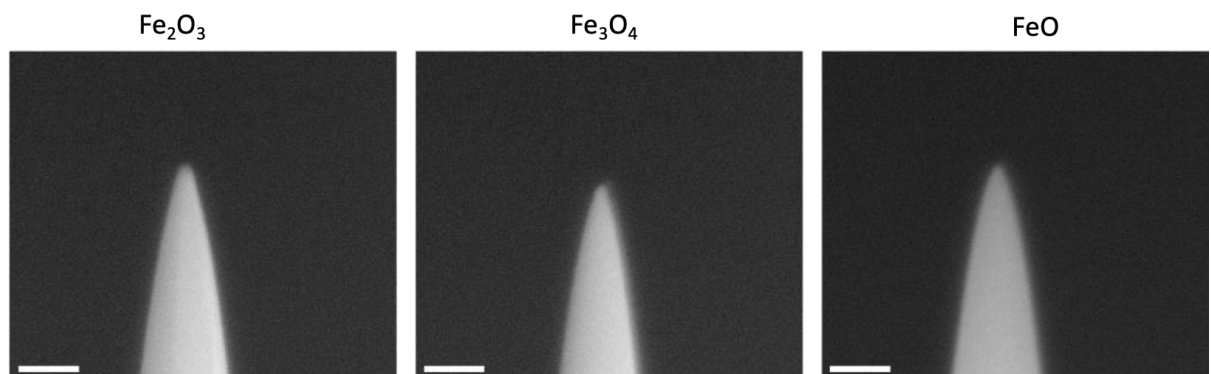

Suppl Figure 1: Scanning electron microscopy images of the tips corresponding to each oxide used for analysis, Showing very similar end radius and shape. The scale bar is 100 nm.

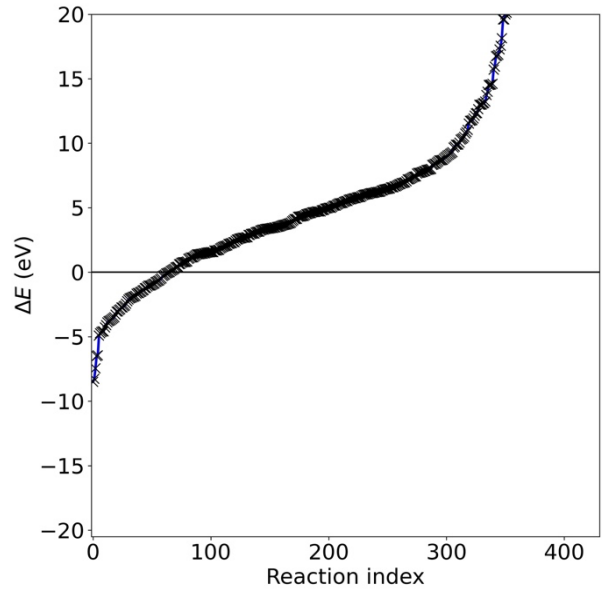

*Suppl Figure 2: Energies of the reactions in increasing order plotted against their reaction index. 369 reactions have an energy predicted by DFT out of which 65 reactions have negative energy (below the horizontal line in the figure).*

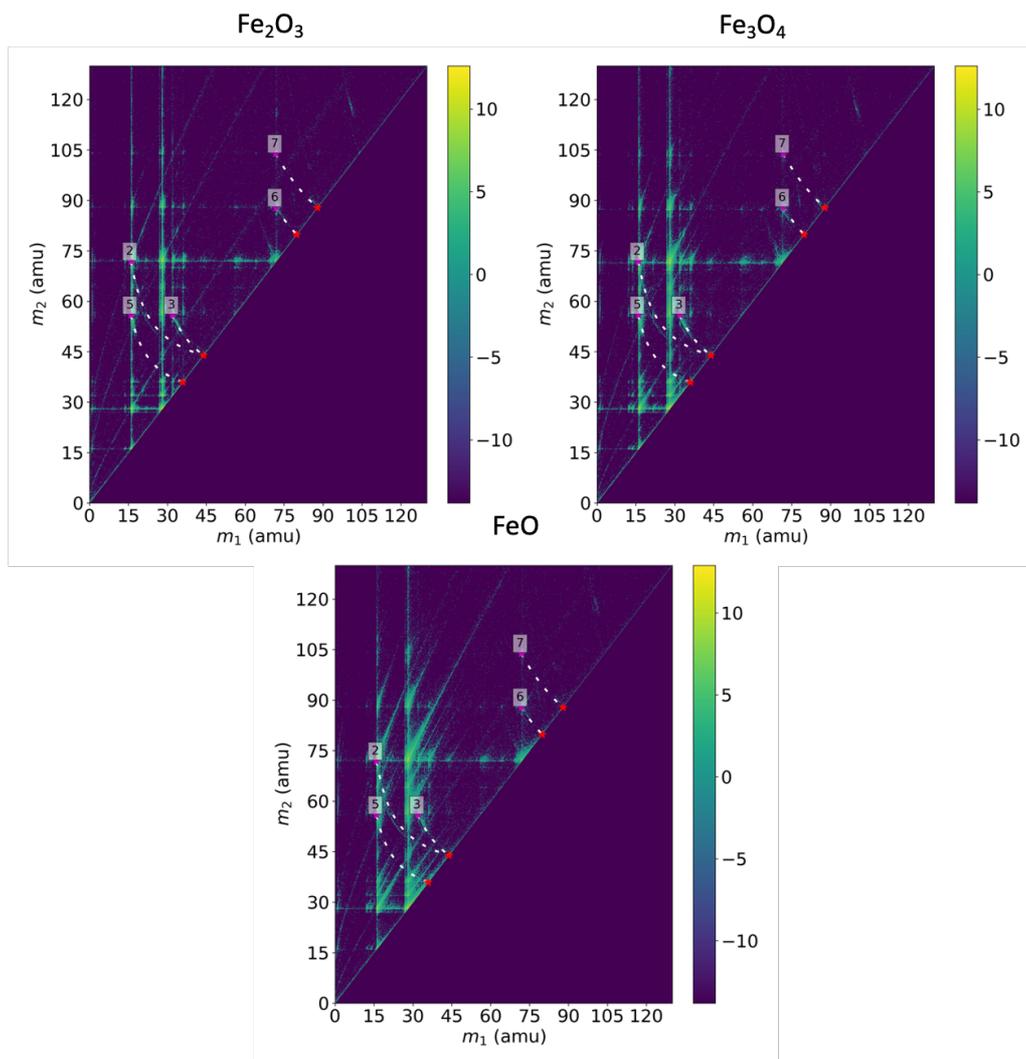

*Suppl Figure 3: Correlation histogram overlaid with detected and DFT predicted reactions involving charged products. The reactions are numbered according to Table 1. The red and blue stars show the parent's and daughters' mass-to-charge ratios.*

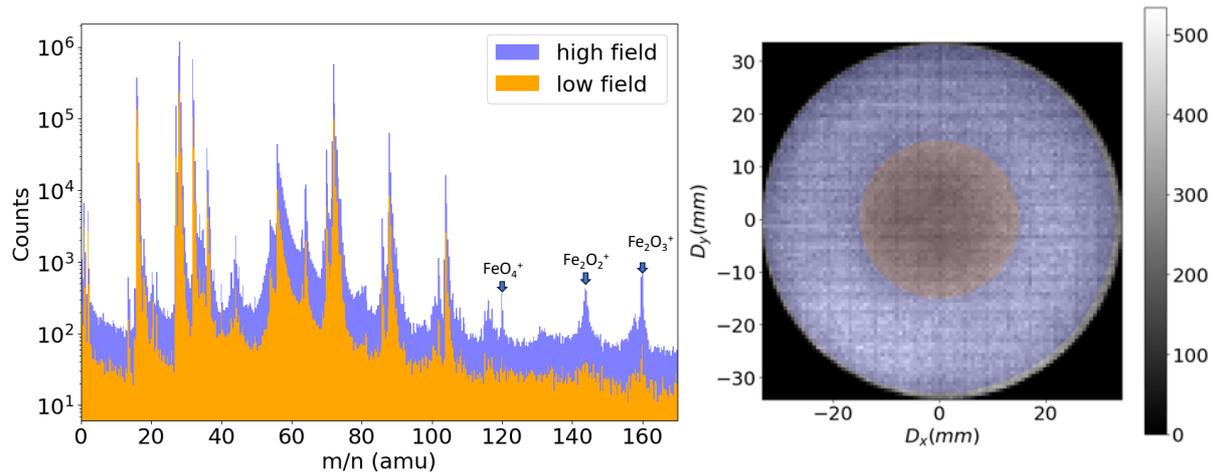

*Suppl Figure 4: Mass spectra corresponding to low (orange, inner ring) and high (blue, outer ring) field regions of the detector. The high field region mass spectrum shows additional peaks corresponding to heaving molecular ion species (arrows).*

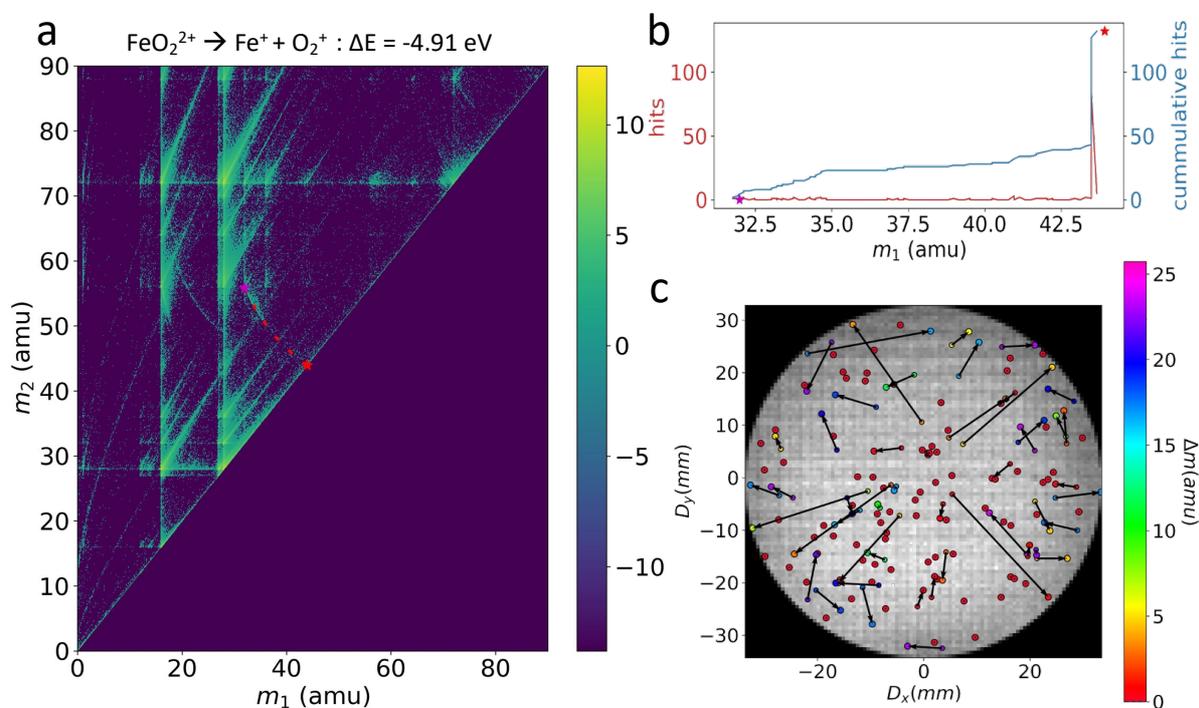

*Suppl Figure 5: a) Correlation histogram of multi-hits with a reaction traced in red for charged daughter species. B) shows the number of hits along the traced reaction track and c) is the impact positions of pairs extracted from the track, which unlike reaction involving neutral daughter species, is more random.*